\documentclass[%
 aip,
 amsmath,amssymb,
 reprint,%
]{revtex4-1}

\usepackage{graphicx}
\usepackage{dcolumn}
\usepackage{bm}

\usepackage[utf8]{inputenc}
\usepackage[T1]{fontenc}
\usepackage{mathptmx}
\usepackage{etoolbox}

\newcommand*{\citen}[1]{%
  \begingroup
    \romannumeral-`\x 
    \setcitestyle{numbers}%
    \cite{#1}%
  \endgroup   
}

\makeatletter
\def\@email#1#2{%
 \endgroup
 \patchcmd{\titleblock@produce}
  {\frontmatter@RRAPformat}
  {\frontmatter@RRAPformat{\produce@RRAP{*#1\href{mailto:#2}{#2}}}\frontmatter@RRAPformat}
  {}{}
}%
\makeatother
\begin{document}

\preprint{AIP/123-QED}

\title[Expanding the Quantum Photonic Toolbox in AlGaAsOI]{Expanding the Quantum Photonic Toolbox in AlGaAsOI}
\author{J. E. Castro}
\affiliation{Electrical and Computer Engineering Department, University of California, Santa Barbara, CA 93106}%
\author{T. J. Steiner}%
\affiliation{Materials Department, University of California, Santa Barbara, CA 93106}
\author{L. Thiel}
\affiliation{Electrical and Computer Engineering Department, University of California, Santa Barbara, CA 93106}%
\author{A. Dinkelacker}
\affiliation{Electrical and Computer Engineering Department, University of California, Santa Barbara, CA 93106}%
\author{C. McDonald}
\affiliation{Electrical and Computer Engineering Department, University of California, Santa Barbara, CA 93106}%
\author{P. Pintus}
\affiliation{Electrical and Computer Engineering Department, University of California, Santa Barbara, CA 93106}%
\author{L. Chang}
\affiliation{Electrical and Computer Engineering Department, University of California, Santa Barbara, CA 93106}%
\author{J. E. Bowers}
\affiliation{Electrical and Computer Engineering Department, University of California, Santa Barbara, CA 93106}%
\affiliation{Materials Department, University of California, Santa Barbara, CA 93106}
\author{G. Moody}
\affiliation{Electrical and Computer Engineering Department, University of California, Santa Barbara, CA 93106}%
 \email{moody@ucsb.edu}

\date{\today}

\begin{abstract}
Aluminum gallium arsenide-on-insulator (AlGaAsOI) exhibits large $\chi^{\left(2\right)}$ and $\chi^{\left(3\right)}$ optical nonlinearities, a wide tunable bandgap, low waveguide propagation loss, and a large thermo-optic coefficient, making it an exciting platform for integrated quantum photonics. With ultrabright sources of quantum light established in AlGaAsOI, the next step is to develop the critical building blocks for chip-scale quantum photonic circuits. Here we expand the quantum photonic toolbox for AlGaAsOI by demonstrating edge couplers, 3-dB splitters, tunable interferometers, and waveguide crossings with performance comparable to or exceeding silicon and silicon-nitride quantum photonic platforms. As a demonstration, we demultiplex photonic qubits through an unbalanced interferometer, paving the route toward ultra-efficient and high-rate chip-scale demonstrations of photonic quantum computation and information applications.
\end{abstract}


\maketitle


\section{\label{sec:intro}Introduction\protect\\ }

\noindent Photonic integrated circuits (PICs) have already shown promise as platforms for quantum information\cite{Moody2020,Vigliar2021,Llewellyn2019,Arrazola2021,Zhao2020,Wang2018}. However, development of PICs for problems of interest in fields such as communication, computing, sensing, and metrology will require circuits with both scale and functional complexity well beyond even the most mature of quantum PICs (QPICs). QPIC functionality of any complexity relies on a small set of components for on-chip manipulation of quantum states of light, including active modulators and switches as well as passive routing components, beamsplitters, and on/off chip couplers \cite{moody20222022}. The largest demonstrations to date using these components and on-chip photon sources have been realized on silicon\cite{Vigliar2021,Ma2017}, whose swift progress as a quantum photonic platform has benefited from its prominence in classical photonics and the complementary metal oxide semiconductor (CMOS) industry. Other demonstrations include material platforms such as  silicon nitride\cite{Zhao2020,Jose2017}, aluminum nitride\cite{Guo2016}, lithium niobate\cite{Zhao20,Boes2018}, and indium phosphide\cite{Kumar2019}; however, no single platform clearly wins out on performance across all metrics, including material absorption, waveguide propagation loss, $\chi^{(2)}$ and $\chi^{(3)}$ nonlinearities, and thermo-optic coefficients for tuning and modulation. Recent improvements in fabrication have opened the possibility of alternative materials with better suitability towards specific functionalities than silicon, such as the aluminum gallium arsenide-on-insulator (AlGaAsOI) platform\cite{Xie2020}.

As a QPIC platform, AlGaAsOI is attractive for its large $\chi^{(2)}$ and $\chi^{(3)}$ nonlinear coefficients, which enable more efficient spontaneous parametric down conversion (SPDC) and spontaneous four wave mixing (SFWM) than silicon, strong modal confinement due to its large refractive index\cite{Semenova2016}, a thermo-optic coefficient comparable to silicon for efficient tuning\cite{Komma2012,Cocorullo1999,Corte2000}, and electro-optic\cite{Walker2019} and piezo-optic\cite{Forsch2020} effects for cryogenic tuning. Additionally, varying the aluminum content of AlGaAs enables bandgap engineering such that two-photon absorption, a major challenge in silicon QPICs, is minimal at telecommunication wavelengths\cite{Adachi1994}. Until quite recently, the utility of the AlGaAsOI platform has been restricted by high waveguide propagation loss\cite{Andronico2014}. Improvements in the fabrication process \cite{Xie2020} have reduced propagation losses to less than 0.2 dB/cm, enabling demonstrations of microring resonators with quality factors above 3 $\times 10^6$. In combination with low material absorption and a large $\chi^{(3)}$ nonlinear coefficient, this has led to record-low Kerr comb generation threshold powers and entangled photon pair sources 1000 times brighter than state-of-the-art in silicon\cite{Chang2020,Steiner2021}. 

\begin{figure*}[!htb]
\centering
\includegraphics[width=\textwidth]{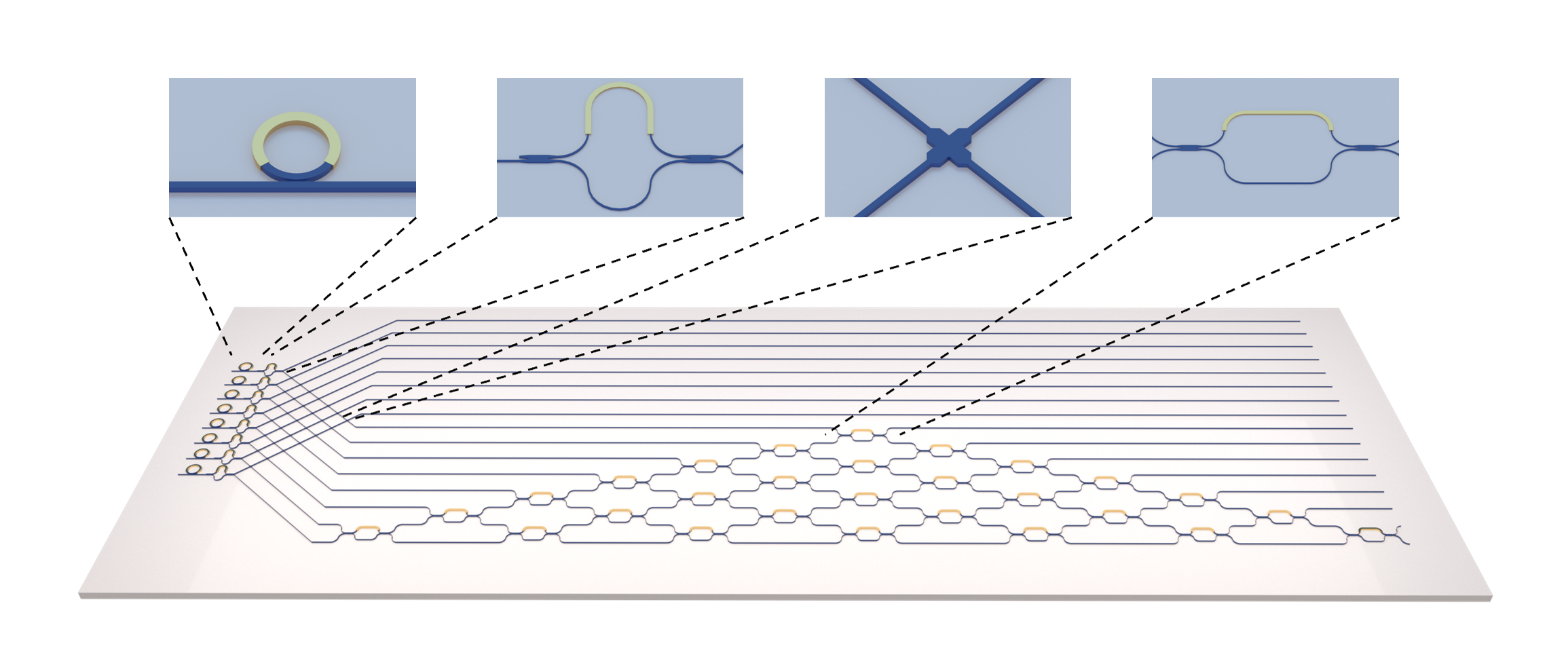}
\vspace{-35pt}
\caption{\label{fig:intro} Illustrative example of a programmable QPIC for on-chip optical computing and communications, shown here as a generic \textit{m}-mode unitary operator for Boson sampling. A general platform comprises: (i) tunable microring resonators or waveguides for quantum light generation; (ii) tunable unbalanced Mach-Zehnder interferometers for filters and qubit demultiplexers; (iii) waveguide crossings for connecting non-nearest neighboring qubits; and (iv) tunable balanced Mach-Zehnder interferometers for linear optical programming.}
\end{figure*}

Because of its novelty, the AlGaAs platform has lacked the same level of individual circuit component development as more established platforms including silicon and silicon nitride. Here, we report on high-performance AlGaAsOI components including edge couplers, waveguide crossings, and tunable interferometers that, along with previously demonstrated ring resonator sources, form the foundation of a functional on-chip platform with potential not only in QPICs but classical PICs as well, including applications in optical communications and spectroscopy\cite{Gaeta2019,Kippenberg2018,Spencer2018}. Interferometers serve many purposes for on-chip optical quantum computing schemes including pump filtering, demultiplexing distinct frequency modes, acting as beamsplitters, and serving as linear optical quantum gates. Following the protocol proposed by Knill, Laflamme, and Milburn \cite{KLM}, scalable linear optical computing can be realized using only linear optical interferometers (reported here), single photon sources (shown in our previous manuscript \cite{Steiner2021}), and single-photon detectors (available commercially and also on-chip\cite{mcdonald2019iii}). Thus, with the design of high quality interferometers and low-loss chip-to-fiber couplers, the necessary components for many linear optical quantum computing protocols have been established. The other component discussed here--waveguide crossings--enables more complex circuit design where interactions between non-next nearest-neighbor qubits is necessary. Since the circuit elements are all fabricated using the same photonic layer, the routing requires waveguide crossings with minimal loss and crosstalk to adjacent channels. The development of these components opens the door for demonstrations of chip-scale QPICs using the AlGaAsOI platform, leveraging the benefits that AlGaAs offers over silicon and other nonlinear optical platforms\cite{ChangCSOI}. The components described below are designed assuming eventual full integration with AlGaAsOI ring resonator entangled-photon pair sources optimized for $\chi^{(3)}$ spontaneous four-wave mixing like that detailed in Ref. [\citen{Steiner2021}]. As an example, Fig. \ref{fig:intro} shows a quantum photonic circuit for chip-scale Boson sampling to illustrate the opportunities that the development of interferometers and waveguide crossings can enable. The individual components are highlighted where they may reside in complex quantum photonic circuitry. In the remainder of this manuscript, we describe the design, simulation, and experimental results of many of the fundamental components to develop a quantum photonic platform.

\vspace{-10pt}

\section{\label{sec:toolbox}Quantum Photonic Toolbox}

\vspace{-5pt}

All of the components were fabricated using an AlGaAs photonic layer grown via molecular-beam epitaxy (MBE). The full fabrication procedure has been detailed previously\cite{Steiner2021,Chang2020}. Briefly, a GaAs chip with a 400-nm-thick AlGaAs photonic layer is bonded onto a 3-$\mu$m-thick thermal SiO$_2$ buffer layer on a Si substrate. After removing the substrate through selective wet etching, the AlGaAs surface is passivated using an 8-nm film of Al$_2$O$_3$ grown via atomic layer deposition (ALD). Deep ultraviolet photolithography is used to pattern photoresist, which is used to etch an SiO$_2$ hardmask and then the AlGaAs photonic layer to define the components. Another ALD deposition of Al$_2$O$_3$ passivates the surface before a 1.0-$\mu$m thick SiO$_2$ cladding layer is deposited. Finally, 10 nm of titanium and 100 nm of platinum are deposited as resistive heaters for thermo-optic tuning.  

\vspace{-10pt}

\subsection{\label{sec:taper}Edge Couplers}

\vspace{-10pt}

Before a fully integrated QPIC is fabricated, it is necessary to utilize either fiber-based or free-space optical testing instruments. Thus, one of the first components to design is an efficient structure to couple light into and out of the photonic chip. Various strategies have been explored for efficient coupling \cite{Marchetti2019}, but many of the methods that achieve ultra-high efficiency require additional fabrication steps, electron-beam lithography, anti-reflection coatings, or a full redesign of the input/output facet structure. For high throughput testing of individual components, much simpler structures can be used; as long as the fiber-to-chip coupling efficiency can be adequately characterized, any undesirable effects due to the input/output coupling can be isolated from the component performance.

\begin{figure}[!b]
\centering
\includegraphics[width=\columnwidth]{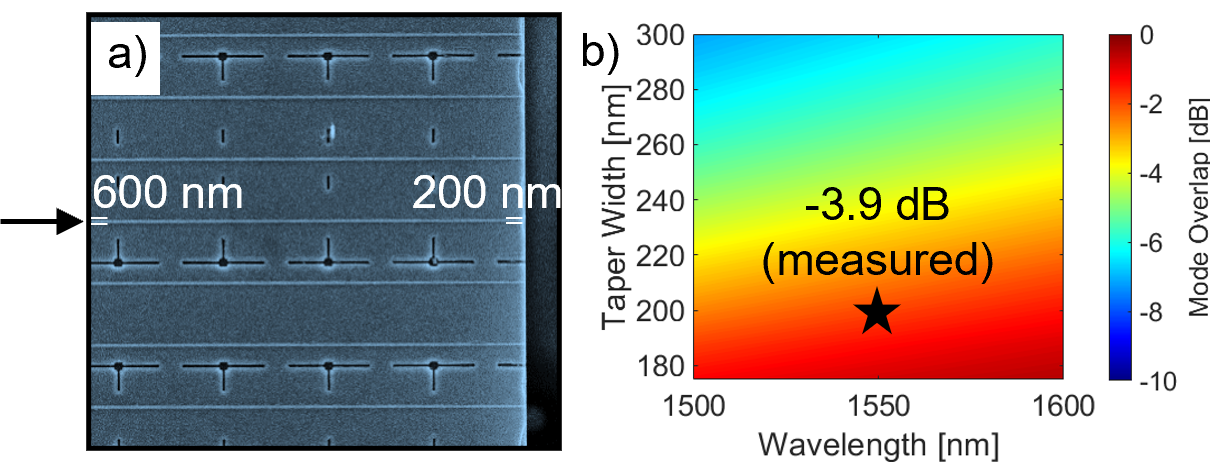}
\caption{\label{fig:edgecoupler}\textbf{a)} Scanning electron microscope (SEM) image of an array of inverse taper edge couplers. \textbf{b)} Simulated mode overlap between a 2.5 $\mu$m Gaussian beam and an AlGaAsOI waveguide for various widths of the input taper and a fixed height of 400 nm. The star indicates an experimental measurement of the coupling loss through a straight waveguide device with 200 nm input/output couplers.}
\end{figure}

There are two main categories of fiber-to-chip couplers: vertical couplers and edge couplers (also called "in-plane" or "butt" couplers, respectively). As the name suggests, vertical couplers accept incoming light from the top of the chip (out of plane) while edge couplers couple light impinging from one of the facets of the photonic chip (in plane). Depending on the desired application or testing design, there may be a benefit for utilizing either type of coupler. Vertical couplers are useful for more compact designs as they do not require waveguides to be routed completely to the edge of the photonic chip\cite{Cheng2020}. Generally, vertical couplers utilize periodic gratings that satisfy the Bragg condition in the waveguide to couple light into an optical fiber oriented almost perpendicularly to the surface \cite{Cheng2020,Taillaert2006}. Edge couplers, on the other hand, can be much simpler to design and less sensitive to fabrication variation. Instead of relying on grating-based structures, edge couplers manipulate the waveguide dimensions to expand the waveguide mode to be closer matched with the mode in the fiber\cite{Mu2020}. The easiest way to achieve this conversion is to taper the waveguide to a narrow tip (called an "inverse taper") where the mode becomes weakly confined and expands closer to the mode size of the fiber. A standard taper, where the waveguide width is expanded at the facets, can also serve as an edge coupler. However, in high-index contrast platforms the design typically has lower efficiencies since the strongly confined mode will remain smaller than the fiber mode. The waveguide becomes capable of supporting multiple modes, and the high-index contrast produces large back reflections\cite{Mu2020}. Other edge coupling strategies include utilizing multiple inverse tapers in a trident or dual-tip design \cite{Yan2007}, polymer-based spot size converters \cite{Roelkens2005}, or multi-layer spot size converters \cite{Fang2010}. Due to the much simpler design, here we report only on inverse taper AlGaAsOI edge couplers. A scanning electron microscope image of several inverse tapers is shown in Fig. \ref{fig:edgecoupler}\textbf{a)}. The facet is on the right side of the image, and the waveguides taper from 600 nm width (on the left) down to 200 nm at the facet. For a more in-depth study of the various types of coupling strategies, readers are referred to Ref. [\citen{Marchetti2019}], which highlights various vertical and edge coupling strategies on the silicon-on-insulator (SOI) platform.

The inverse taper design reduces the confinement of the waveguide mode, increasing its effective modal area and decreasing its effective index of refraction. This allows for moderately high simulated coupling efficiencies (losses $<3$ dB) to a (typically) lensed fiber aligned with the waveguide facet. The overall coupling loss is determined by effects such as reflection at the chip facet (due to refractive index mismatch), fiber-to-waveguide mode mismatch, and mode-conversion within the waveguide taper. 

To determine the optimal design for an inverse taper edge coupler, the dimensions of the waveguide taper were varied and simulated using Lumerical Mode software. Here we only show the results for inverse tapers designed for the fundamental transverse electric (TE) mode because the components shown in the rest of this article are designed to operate with TE polarized light. Similar calculations can be made for the transverse magnetic (TM) mode. For a given Gaussian beam and waveguide geometry, the power overlap between the waveguide mode and fiber mode is calculated to estimate an upper bound on coupling efficiency and determine the optimum waveguide dimensions. This calculation does not include loss due to mode conversion or reflection at the interface. Figure \ref{fig:edgecoupler}\textbf{b)} illustrates the simulated mode overlap between a Gaussian beam with a mode field diameter of 2.5 $\mu m$ (which matches the mode field diameter of commercially available lensed fibers) and a 400-nm thick AlGaAsOI waveguide with various taper widths. Narrow taper widths enlarge the waveguide mode to be nearly mode-matched with the incoming fiber mode, but the weak confinement of these narrow waveguides typically comes with additional loss as the light propagates through the narrow taper back to a waveguide width of $\ge$400 nm for the components. The fabrication of sub-200-nm features is challenging using the standard photolithography process, so we limit our taper designs to 200 nm or larger. Along with the simulated data, Fig. \ref{fig:edgecoupler}\textbf{b)} also shows a measured value for the coupling loss for a 200 nm edge coupler at a wavelength of 1550 nm. The measured value was collected by sending 10 dBm (10 mW) of light into a straight waveguide with 200 nm inverse tapers on the input and output facet. The collected power through the waveguide was 2.2 dBm, indicating an approximate loss of 3.9 dB/facet (the waveguide propagation loss is $<1$ dB/cm and the waveguide is less than 2 mm, so the contributions of propagation loss are ignored in this measurement). The measured loss is larger than the simulated mode overlap, which is expected because the measurements also include reflections and mode conversion loss in the taper. The simulated mode overlap acts as an upper bound for the efficiency of the inverse taper. The 3.9 dB/facet of coupling loss in the AlGaAsOI platform is slightly larger than the sub-3 dB coupling loss expected from standard SOI inverse taper edge coupler designs \cite{Mu2020}. The use of narrower taper widths (as shown in Fig. \ref{fig:edgecoupler}\textbf{b)}) or an anti-reflection coating will improve the coupling efficiency further, but ease of fabrication and reliability are prioritized, so for our initial devices, 200 nm inverse tapers are utilized.

\subsection{\label{sec:crosser}Waveguide Crossings}

Several methods have been explored for creating low-loss waveguide crossings including vertical coupling into polymer strip waveguides \cite{Tsarev2011}, multi-planar crossings\cite{Chiles2017}, multimode interference-based crossings\cite{WChang2020,Bogaerts2007,Wu2020}, and subwavelength gratings\cite{Bock2010}. Many of these methods involve additional fabrication steps that can introduce excess loss and system design and fabrication challenges. A basic approach for a waveguide crossing relies on tapering an input single-mode waveguide section into a larger waveguide cross section that can support higher-order modes and relies on the beating between the fundamental mode and the higher-order mode to create an electric field maxima that is centered in the waveguide at the crossing location. By focusing the mode into the center of the wide waveguide, evanescent coupling to the perpendicular waveguide is minimized. This design can be completed with a basic linear taper (which will be referred to as a "simple crossing"), or a more complex structure. Here, we consider simulations of both simple and inverse-design crossings and report results from an inverse design approach (which will be referred to as a "13-width crossing") that utilizes 13 different widths in a parabolic taper that requires no additional fabrication steps and maintains low-loss and high-isolation transmission. This second design utilizes a swarm optimization protocol such that the optical mode is transmitted with minimal coupling to the crossed waveguide. 

\begin{figure}[!b]
\centering
\includegraphics[width=\columnwidth]{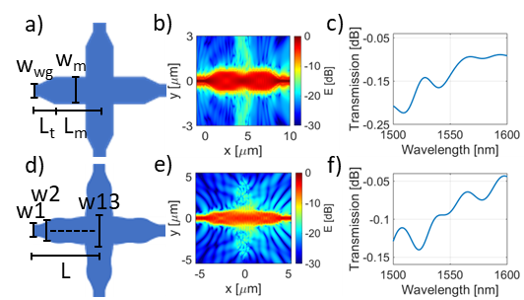}
\caption{\label{fig:crossing}\textbf{a)} Schematic image of the simple waveguide crossing design. The relevant design variables are indicated on the image.  \textbf{b)} Simulated mode profile for the optimized simple crossing design. The beating between the higher order modes and the fundamental mode results in a maximum electric field intensity in the center of the waveguide at the crossing, which minimizes the coupling to the vertical waveguide. \textbf{c)} Simulated transmission through the simple waveguide crossing as a function of the wavelength. \textbf{d)} Schematic image of the 13-width crossing design where a swarm optimization varies the width of the parabolic taper at 13 locations and minimizes the transmission loss. \textbf{e)} Simulated mode profile of the optimized 13-width crossing. \textbf{f)} Simulated transmission through the 13-width crossing design as a function of wavelength. }  
\end{figure}

The simple crossing design is illustrated in Fig. \ref{fig:crossing}\textbf{a)} and uses the beating between the fundamental and higher-order mode to create a confined optical mode centered at the location of the crossing. The beat length, \textit{$L_\pi$}, is defined as $L_\pi=\pi/(\beta_0-\beta_1)$ where $\beta_{(0,1)}$ are the propagation constants of the fundamental and first-order waveguide modes, respectively. For a 1.5 $\mu$m multimode waveguide width (\textit{$w_m$}), the fundamental and first-order TE modes have effective indices of approximately 3.00 and 2.87, respectively, at a wavelength of 1550 nm. Using these effective indices, the beat length is calculated as 5.95 $\mu$m. Finite difference time domain (FDTD) simulations were utilized to account for the transition region between the single-mode waveguide and the wider, multimode waveguide as well as allow for a larger bandwidth crossing to be designed where the average loss across a 100 nm bandwidth is utilized instead of maximizing at a single wavelength. The simulated mode profile for the simple crossing design and the simulated transmission through the device is shown in Fig. \ref{fig:crossing}\textbf{b,c)}, respectively. From the mode profile simulation, it is clear that the beating between the fundamental and higher order mode create a local maxima at the location of the crossing that minimizes the optical mode scattered into the perpendicularly oriented waveguide. The simulated loss through this structure is 0.15 dB at a wavelength of 1550 nm.

Figure \ref{fig:crossing}\textbf{d)} shows the 13-width waveguide crossing with critical dimensions depicted. This design utilizes a swarm optimization protocol in an FDTD solver to optimize the transmission through the crossing by allowing the width to vary at 13 equally spaced sections along the taper. A parabolic interpolation between the 13 widths ensures a smooth transition between the various widths. The 13-width crossing design was also optimized for a bandwidth of 100 nm to maintain low-loss performance of the crossing across a broad bandwidth, which will be compatible with broadband entangled photon pair generation in quantum photonic circuits. Lower loss structures can be made when optimizing for a smaller bandwidth. Starting with an input waveguide width of 400 nm and total crossing length (\textit{L}) of 9 $\mu$m, the optimizer was allowed to vary the widths \textit{w2-w13} between 200 nm and 2000 nm. Figure \ref{fig:crossing}\textbf{e)} shows the electric field profile for the optimal crossing design at a wavelength of 1550 nm, and Fig. \ref{fig:crossing}\textbf{f)} plots the simulated transmission through the waveguide crossing as a function of the input wavelength. This crossing design has a simulated loss of approximately 0.1 dB at a wavelength of 1550 nm. Since the loss of the 13-width crossing design is smaller than the simple crossing design, the 13-width crossing was fabricated and tested initially.


\begin{figure}[!t]
\centering
\includegraphics[width=1.0\columnwidth]{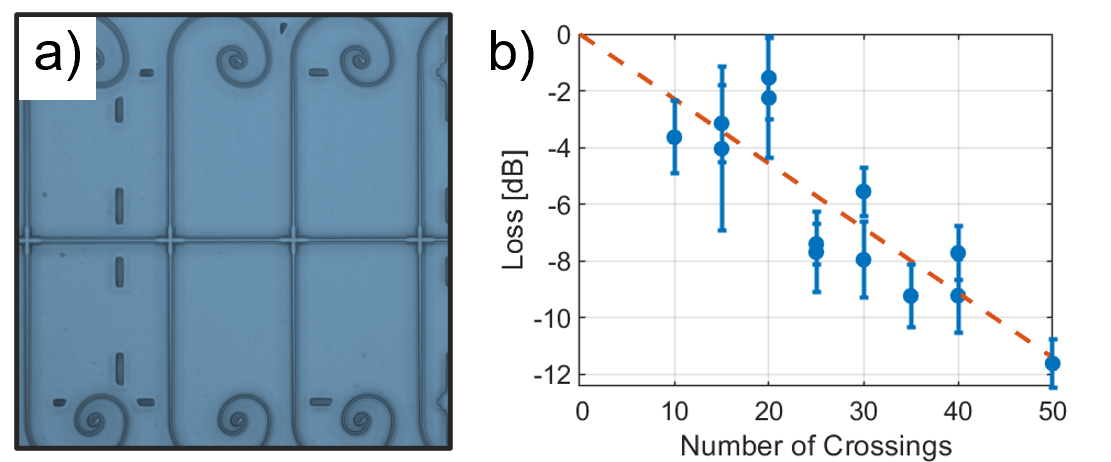}
\caption{ \textbf{a)} Optical image of cascaded waveguide crossings for loss characterization. \textbf{b)} Normalized waveguide crossing transmission loss at a wavelength of 1550 nm using the cutback method. The orange trend line indicates an insertion loss of $0.23$ dB/crossing.}
\label{fig:1}
\end{figure}

With the simulated waveguide crossing loss on the order of 0.1 dB, the cutback method\cite{Bogaerts2007} is used to measure the loss per crossing to remove coupling-dependent loss and reduce detector sensitivity limitations. For the 13-width crossing, waveguides with between 10 and 50 crossings were fabricated, and the loss through each line of crossings was measured across eight trials with complete re-alignment of the input and output fibers for each trial to remove any systematic variations due to coupling loss. Figure \ref{fig:1}\textbf{a)} shows a microscope image of a few of the waveguide crossings in one of the lines. The vertical waveguide channels are terminated with tapered waveguides in a spiral geometry to prevent back-reflections into the crossing. The horizontal spacing of the crossings is varied randomly between 25 $\mu$m and 35 $\mu$m to avoid photonic cavity effects. Using the cutback method, the transmission through the crossings was measured at a wavelength of 1550 nm, and the results are shown in Fig. \ref{fig:1}\textbf{b)}. The dashed line indicates a linear fit of the loss as a function of the number of crossings, providing an estimated loss of $0.23$ dB/crossing. The error bars on the data points indicate the standard deviation of the eight independent measurement trials.

These results for the 13-width waveguide crossing (Fig. \ref{fig:crossing}\textbf{b}) indicate that the fabricated crossings have slightly higher loss than the simulated loss at a wavelength of 1550 nm. This additional loss is likely due to fabrication variations in the widths along the device; the inverse design is more sensitive to fabrication variation than the use of a simple waveguide crossing. The measured $0.23$ dB of loss for the AlGaAsOI 13-width crossing is comparable to the $0.2$ dB of insertion loss reported from a genetic algorithm-designed SOI waveguide crossing \cite{Sanchis2009} and less than the loss of $0.3$ dB from silicon nitride waveguide crossings \cite{Yang2019}. Other manuscripts report $\le0.1$ dB of insertion loss for elliptical tapers \cite{Fukazawa2004} and even on the order of $0.02$ dB for sub-wavelength grating-based structures\cite{Zhang2013}. Ref. [\citen{Wu2020}] compares various results of waveguide crossing on the SOI platform.

\subsection{\label{sec:mmi}3 dB Couplers \protect\\Multimode Interferometers and Directional Couplers}

A standard building block in both classical and quantum PICs is a 3-dB coupler. In QPICs, 3-dB couplers are utilized as their classical counterparts to distribute light evenly between two waveguides, to interfere single photons, and to serve as a component for tunable Mach-Zehnder Interferometers (MZIs) for programmable PICs. This places strict requirements on the devices, such as low loss for potential scalability, a large bandwidth to support broadband quantum light generation, and precise splitting ratios to maximize the extinction ratio and minimize cross-talk in MZIs. We explore two designs for creating on-chip 3-dB couplers: multimode interferometers (MMIs) and directional couplers (DCs). MMIs are based on the self-imaging principle, similar to the aforementioned simple waveguide crossings; however, unlike in the waveguide crossing design, the beat length between the two modes, \textit{$L_\pi$}, is used to calculate the core length necessary to achieve a splitting ratio as close to 3-dB as possible. The second coupler design based on DCs uses the overlap of evanescent modes between two neighboring waveguides, allowing the mode to fully couple into the adjacent waveguide. The full crossover length relies on the difference in refractive index between the even and odd supermodes created when two waveguides are in close proximity. DCs are straightforward couplers to design and are capable of any splitting ratio by adjusting the coupling length, but they are also more susceptible to fabrication imprecision and errors compared to MMIs.


\begin{figure}[!tbh]
\centering
\includegraphics[width=1.0\columnwidth]{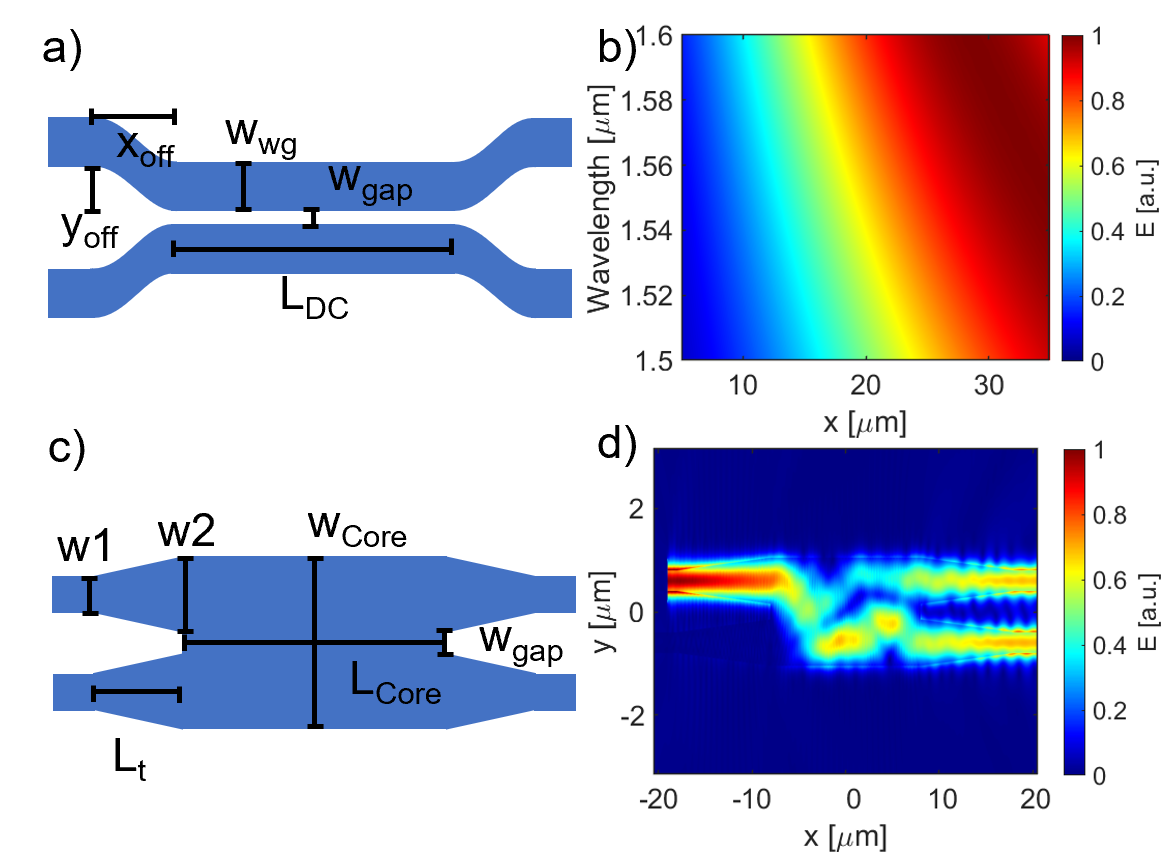}
\caption{\label{fig:coupler} \textbf{a)} Schematic illustration of a directional coupler with the relevant design parameters. \textbf{b)} Simulated transmission of various directional coupling lengths to determine the appropriate coupling length for a 3-dB directional coupler. \textbf{c)} Schematic illustration of a multimode interferometer with the relevant design parameters. \textbf{d)} Simulated mode profile of an MMI depicting 3-dB coupling behavior with a length near 16.0 $\mu$m. }
\end{figure}

Figure \ref{fig:coupler} depicts two MMI and DC designs and results from FDTD simulations. For the MMIs, a core width of 2.1 $\mu$m was selected. Because the self-imaging length scales with the MMI core width, a narrow width was chosen to reduce the component footprint. Symmetric input and output tapers expand the mode from a waveguide width of 0.4 $\mu$m to 0.9 $\mu$m nearest to the core. A 0.3 $\mu$m separation leaves no excess core width beyond the dimensions of the tapers in an effort to reduce Fabry-Perot effects due to reflections. The core length design began by first calculating the beat length and multiplying it by a factor of 1.5, resulting in $L_\pi=18.75$ $ \mu$m. The full device was then simulated using FDTD, and the electric field profile is shown in Fig.\ref{fig:coupler}\textbf{b)}. With a combination of the calculated beat length and FDTD simulations, a core length of 17.2 $\mu$m was chosen. The 1.55 $\mu$m difference between both methods is due to the input taper expanding the mode prior to reaching the MMI core not being considered during the beat-length calculation. 

Unlike the MMIs, the DC design utilizes the same waveguide width, 0.4 $\mu$m, across the entire device. Symmetric sine bend waveguides on the input and outputs with transverse displacements of 1.0 $\mu$m and 8.0 $\mu$m enable light to propagate near the coupling region. The minimum radius of these sine bends is kept to 20 $\mu$m to reduce bending loss. The separation between the waveguides in the coupling region where the evanescent modal overlap occurs is 0.3 $\mu$m. The coupling length for the full transfer of light from one waveguide to the other was first calculated with $L=\frac{\lambda}{2(n_0-n_1)}$ to give an estimate of the full crossover length of the mode, where $n_{(0,1)}$ is the effective indices of the even and odd supermodes that exist when the two waveguides are brought in close proximity. The finite difference eigenmode (FDE) results for the full crossover length is 48.47 $\mu$m. Thus for a 3-dB coupler, $L=24.23$ $\mu$m.  A sweep of the coupling region using FDTD simulations of the full DC structure is depicted in Fig. \ref{fig:coupler}\textbf{d)}. The results of this simulation suggest an optimal 3-dB coupling length of 17.0 $\mu$m. The difference between the two values is due to extra coupling effects in the sine bends. From the MZI measurements discussed in the next section, we can extract the performance of the couplers.

\subsection{\label{sec:mzi}Mach-Zehnder Interferometers (MMI and DC)}

\begin{figure}[thb!]
\centering
\includegraphics[width=1.0\columnwidth]{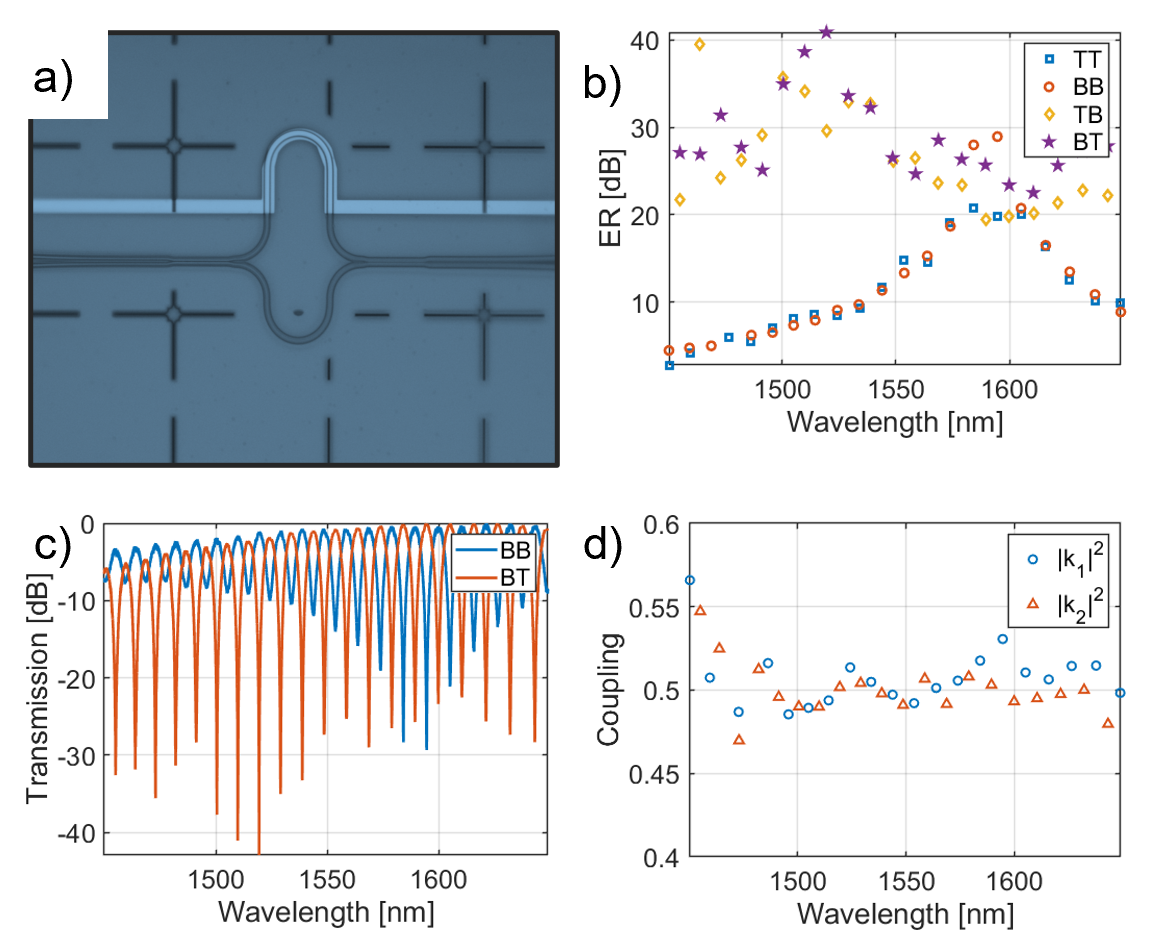}
\caption{\label{fig:MZI}   \textbf{a)} Optical image of an MZI (DC) with a 30 $\mu$m path imbalance. \textbf{b)} Extinction ratios of the different input and output combinations in a 10.23 nm FSR MZI (DC). \textbf{c)} Transmission spectra of the bottom input, bottom output and bottom input, top output port combinations of an MZI (DC). \textbf{d)} Extracted coupling values of each directional coupler in the MZI (DC).}
\end{figure}

Tunable MZIs are a key component in QPICs by playing an important role for numerous functions, including as reconfigurable postselected entangling gates (R-PEGs)\cite{Adcock2018}, demultiplexers\cite{Horst2013}, variable beamsplitters\cite{Wilkes2016}, filters\cite{Piekarek2016}, and single photon quantum logic gates\cite{Adcock2019}. In an MZI, a 3-dB coupler splits light evenly into two different paths that may be equal (balanced MZI) or unequal (unbalanced MZI) in length, which then recombine with another coupler. Here, we focus on two variations of thermo-optically tunable unbalanced MZIs employing both DCs and MMIs. These devices were designed using the transfer matrix method\cite{Tran2016}, where each component of the MZI can be represented by a matrix, two equivalent matrices for the 3-dB couplers, and a standalone matrix representing the path imbalance. Since many MZIs are required for a complete QPIC, the loss across each device must be minimized. Each coupler also should exhibit as close to a 3-dB splitting ratio as possible to achieve a maximum extinction ratio (ER), defined here as the power ratio of neighboring MZI fringes in the transmission spectrum.

Figure \ref{fig:MZI}\textbf{a)} shows an optical image of an MZI utilizing DCs as couplers with a 30 $\mu$m path imbalance on the top arm with the metal thermal tuner above the 1 $\mu$m thick cladding to sweep and control the MZI phase. One advantage of thermo-optic tuning with AlGaAs is its inherent large thermo-optic coefficient, which, for an MZI with a 60 um path imbalance and a 10.28 nm free spectral range (FSR), allows for a full 2$\pi$ phase sweep with 20 mW/$\pi$ efficiency, which is 10 (0.6) times more efficient than silicon nitride\cite{Lee2022} (silicon\cite{Lee2017}). The transmission spectrum of a 60 $\mu$m path imbalance MZI with DC couplers is shown in Fig. \ref{fig:MZI}\textbf{b)} for two different input/output configurations. The legend indicates which port light was coupled into and collected from, respectively, where \textit{T} represents \textit{top} and \textit{B} represents \textit{bottom}. The ER of all four ports are graphed in Fig. \ref{fig:MZI}\textbf{c)}. We typically observe an ER above 10 dB across $\ge100$ nm bandwidth for through ports and $\ge200$ nm for cross ports, comparable to silicon MZIs\cite{Lee2019,Piekarek2016}. With these ER measurements versus wavelength, the true coupling coefficient $\kappa$ of each DC can be extracted\cite{Tran2016}, as shown in Fig. \ref{fig:MZI}\textbf{d)}. The DC couplers exhibit an average coupling coefficient of 0.504$\pm$0.034 across a 100 nm bandwidth centered at 1570 nm.



\section{\label{sec:demux}Qubit Demultiplexing}

\begin{figure}[!bht]
\centering
\includegraphics[width=1.0\columnwidth]{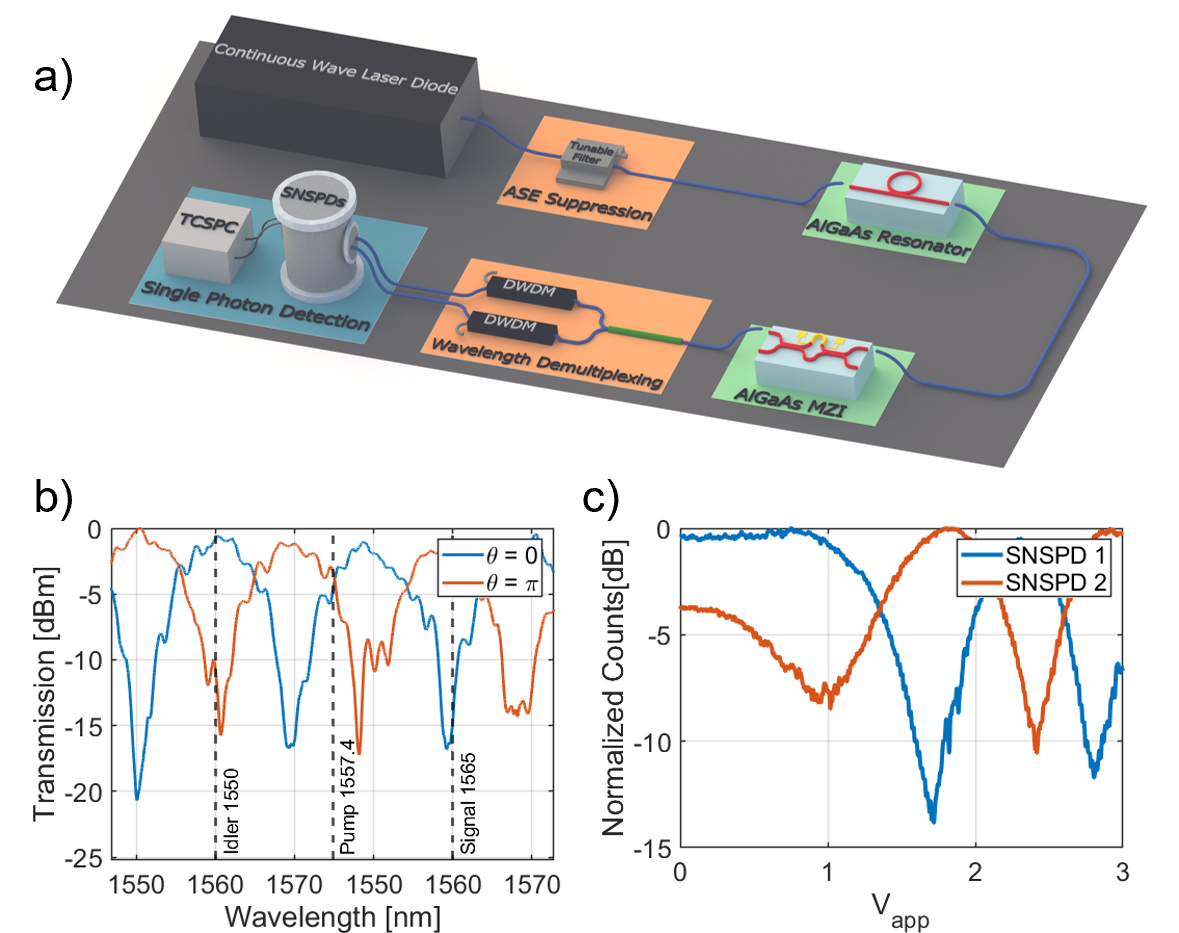}
\caption{ \label{fig:demux} \textbf{a)} Experimental schematic of qubit demultiplexing utilizing a AlGaAs ring resonator source and a 10.23 nm FSR MZI (MMI). \textbf{b)} Transmission spectra of MZI tuned to relative zero and $\pi$ along with the pump, signal, and idler wavelengths. \textbf{c)} Normalized counts on the SNSPDs after demultiplexing.}
\end{figure}

To benchmark our platform, we use a tunable MZI chip to demultiplex signal and idler photons generated from SFWM in a separate AlGaAsOI microring resonator chip previously reported\cite{Steiner2021}. The experimental setup schematic is shown in Figure \ref{fig:demux}\textbf{a)}. The microring resonator temperature is stabilized with a thermo-electric cooler and control electronics. A tunable, narrow-linewidth continuous wave (cw) laser is tuned into resonance with a microresonator mode near 1557 nm. Tunable etalon-based bandpass filters are tuned to the pump wavelength and placed immediately after the laser to suppress amplified spontaneous emission at the wavelengths of the signal and idler modes, which helps improve the coincidence-to-accidental ratio in our experiments. The laser is polarized along the TE mode of the microresonator chip, and light is coupled onto the chip with high-numerical aperture lensed fiber. Pump light is coupled into the microring resonator using a critically coupled pulley coupler, where time-energy entangled signal and idler photon pairs are generated. The signal, idler, and pump photons are collected with a lensed fiber and then routed with polarization maintaining fiber to the MZI chip. When the phase of the MZI is properly tuned, it demultiplexes the signal and idler photons into separate waveguides. Light from one output port is coupled off chip using a lensed polarization-maintaining single-mode fiber. The output is then coupled to a 3-dB fiber beamsplitter, the output of which is demultiplexed with low-loss narrow-band filters and sent to superconducting nanowire single-photon detectors (SNSPDs) for coincidence detection.

We chose to use an MMI-based MZI with a 60 $\mu$m path imbalance for these experiments, despite the fact that its maximum ER is shifted away from 1550 nm due to the MMI design, because the FSR of this design most closely matches that of our microring resonator entangled-photon pair source. While the ER of this MZI is lower than that measured for the DC-based MZI, we show below that the performance of our demultiplexing experiment is at the maximum capability of our MZI based on its ER, demonstrating the promise of this platform for future all-on-chip demultiplexing. Figure \ref{fig:demux}\textbf{b)} shows the transmission spectrum of the MZI chip when coupling into and out of the top ports for two different thermo-optic heater voltages of 1 V and 1.8 V, which correspond to MZI phase 0 and $\pi$ radians, respectively, with an extinction ratio of 15 dB near the signal and idler wavelengths. As the phase is swept from 0 to $\pi$, the transmission at a single wavelength sweeps from maximum to minimum. Vertical lines in the plot depict the wavelengths of the pump (1557.4 nm), signal (1565 nm), and idler (1550 nm) photons from our microring resonator. We next couple the entangled-photon pairs into the MZI for demultiplexing. Normalized counts from the two SNSPDs are shown in Fig. \ref{fig:demux}\textbf{c)} as a function of applied MZI voltage. We measure an extinction ratio of $\sim$14 dB near 1.65 V, which is comparable to the measured MZI extinction ratio shown in Fig. \ref{fig:demux}\textbf{b)}. While these results are encouraging, we expect further improvements demultiplexing by improving the MMI splitting ratios to increase the demultiplexing ER and by designing an MZI with twice the FSR of the source.

\section{\label{sec:concl}Conclusion}

Here we demonstrate many of the fundamental components necessary to develop fully integrated quantum photonic circuits on AlGaAsOI. With high-quality entangled photon pair sources\cite{Steiner2021} and the efficient edge couplers, 3 dB splitters, waveguide crossings, and MZIs demonstrated in this manuscript, a plethora of application-oriented integrated quantum circuits become available. Demonstrations of chip-to-chip quantum teleportation\cite{Llewellyn2019}, multi-photon quantum information processing \cite{Adcock2019}, and other large-scale quantum photonic circuits have already been realized on the SOI platform\cite{Moody2020,Lu2021}. The benefits of the AlGaAsOI platform should enable more efficient demonstrations of these circuits at significantly lower optical pump power, reducing the required time to collect useful data and allowing for larger-scale circuits to be created. A summary of a few of the components discussed in this report is shown in Table \ref{tab:Table1} along with the performances of comparable components made on the SOI and Si$_3$N$_4$ platforms, which are also commonly used for quantum photonic circuits. It is important to note that the selected device performances were for Si and Si$_3$N$_4$ components that follow similar designs to the AlGaAsOI components that are relevant and routinely used for QPICs. For example, sub-edge couplers with 0.35 dB of loss have been fabricated using silicon with silicon nitride, but these were achieved with multiple layers\cite{Wang2019}. Here we compare similar component designs across the three platforms--using only a single photonic layer and standard photolithography to fabricate the devices. Overall, the AlGaAsOI components have similar or better performance to their SOI and Si$_3$N$_4$ counterparts, indicating that the transition to fully AlGaAsOI-based photonic circuits will have little to no degradation in performance compared to the current state-of-the-art platforms. Although the components detailed in this work were fabricated using a 22 mm by 24 mm bonded AlGaAs chip, wafer-scale bonding with compound-semiconductor-on-insulator is possible\cite{Stanton2020,ChangCSOI}, enabling larger circuits to be created in the near future.

\begin{table}[]
    \caption{Table comparing the AlGaAsOI platform with SOI and Si$_3$N$_4$ designed for integrated quantum photonics.}
    \label{tab:Table1}
    \begin{ruledtabular}
    \centering
    \begin{tabular}{c|c|c|c}
           &AlGaAsOI  &SOI & Si$_3$N$_4$\\
           & (this work)& &\\
         \hline
         \rule{0pt}{4ex}  Inverse Taper  & $3.9$ dB & $<3$ dB\cite{Mu2020} & $2-3$ dB\cite{Ramelow2015}\\
         Coupling Loss & &\\
         \rule{0pt}{4ex}  Waveguide Crossing & $0.23$ dB & $0.2$ dB\cite{Sanchis2009} & $0.3$ dB\cite{Yang2019}\\
         Loss & & &\\
         \rule{0pt}{4ex}  MZI Extinction & $> 30$ dB & $> 30$ dB\cite{Wang2016} & $> 40$ dB\cite{Rao2021}\\
         Ratio & & &\\
         \rule{0pt}{4ex}  MZI Bandwidth & $200$ nm Cross & $> 40$ nm\cite{Lee2019} &$180$ nm\cite{Rao2021} \\
         ($>10$ dB ER) & $90$ nm Through & & \\
         \rule{0pt}{4ex} MZI Heater & 20 mW/$\pi$ & $12$ mW/$\pi$\cite{Lee2017} & $200$ mW/$\pi$\cite{Lee2022} \\
         Efficiency & (10.2 nm FSR) & (5.8 nm FSR) & (NA) \\
    \end{tabular}
    \end{ruledtabular}
\end{table}

\begin{acknowledgments}
This work was supported by the NSF Quantum Foundry through Q-AMASE-i program Award No. DMR-1906325, AFOSR YIP Award No. FA9550-20-1-0150, and NSF Award CAREER-2045246. We also gratefully acknowledge support from the Cisco Research University Gift Program. T.J.S. and J.E.C. contributed equally to this work.
\end{acknowledgments}

\section*{Data Availability Statement}

The data that support the findings in this study are available from the corresponding author upon reasonable request.

\bibliography{bibliography}
\end{document}